# Integrator Anti-Windup Design for Servo-Controllers with Position Constraints

Eugene Lavretsky

*Abstract* – **A control design modification to prevent integrator windup for position saturated servo-controllers is introduced. The design is based on the formalism of Control Barrier Functions and represents an anti-windup integrator modification for position-limited servo-controllers. The method is applicable to Linear Time Invariant Multi-Input-Multi-Output open-loop stable continuous time systems. A flight control application example of the developed anti-windup control solution is discussed.**

## 1. Introduction and Problem Formulation

Consider the Multi-Input-Multi-Output (MIMO) Linear Time Invariant (LTI) dynamical system,

$$\begin{cases} \dot{x}_p = A_p x_p + B_p u \\ y_{reg} = C_{p\,reg} x_p + D_{p\,reg} u \end{cases} \quad (1.1)$$

where $x_p \in R^{n_p}$ is the $n_p$-dimensional state vector, $u \in R^m$ is the $m$-dimensional control input, $y_{reg} \in R^m$ is the system $m$-dimensional vector of regulated outputs and the system matrices $(A_p, B_p, C_{p\,reg}, D_{p\,reg})$ are of the corresponding dimensions. It is further assumed that $A_p$ is Hurwitz, the matrix pair $(A_p, B_p)$ is controllable and that the entire state vector $x_p$ is accessible for control design, as the system output measurement.

Augmenting (1.1) with the integrated tracking error dynamics,

$$\dot{e}_{yI} = \underbrace{y_{reg} - y_{cmd}}_{e_y} \quad (1.2)$$

gives the $n = (n_p + m)$-dimensional extended open-loop system.

$$\underbrace{\begin{pmatrix} \dot{e}_{yI} \\ \dot{x}_p \end{pmatrix}}_{\dot{x}} = \underbrace{\begin{pmatrix} 0_{m \times m} & C_{p\,reg} \\ 0_{n_p \times m} & A_p \end{pmatrix}}_{A} \underbrace{\begin{pmatrix} e_{yI} \\ x_p \end{pmatrix}}_{x} + \underbrace{\begin{pmatrix} D_{p\,reg} \\ B_p \end{pmatrix}}_{B} u + \underbrace{\begin{pmatrix} -I_m \\ 0 \end{pmatrix}}_{B_{cmd}} y_{cmd} \qquad (1.3)$$

It can be proven that (1.3) is controllable if and only if the matrix $\begin{pmatrix} A_p & B_p \\ C_{p\,reg} & D_{p\,reg} \end{pmatrix}$ is nonsingular, which in turn is equivalent to require that the original system (1.1) with the regulated output $y_{reg}$ has no transmission zeros at the origin.

Of interest is the servo-control problem to calculate a baseline state feedback control policy,

$$u = - \underbrace{K_x}_{(K_I \quad K_P)} \underbrace{x}_{\begin{pmatrix} e_{yI} \\ x_p \end{pmatrix}} = -K_I\, e_{yI} - K_P\, x_p \qquad (1.4)$$

such that the system regulated output $y_{reg}$ tracks external bounded commands $y_{cmd}$. In (1.6), $K_I \in R^{m \times m}$ and $K_P \in R^{m \times n_p}$ denote the integral and the proportional feedback gain matrices. It is well-known that a state feedback servo-controller such as (1.4) can be found via the pole placement method or using the Linear Quadratic Regulator (LQR) design [6].

Suppose that in addition to tracking external commands, the servo-controller (1.4) must satisfy min/max position constraints, defined component-wise.

$$u_i^{\min} \leq u_i \leq u_i^{\min}, \quad 1 \leq i \leq m \qquad (1.5)$$

In that case, the servo-controller (1.4) must operate in the presence of position saturation limits,

$$\dot{e}_{yI} = y_{reg} - y_{cmd}$$
$$u = \mathrm{sat}_{u^{\min}}^{u^{\max}}\Big( \underbrace{-K_I\, e_{yI}}_{u_I} \underbrace{-K_P\, x_p}_{u_P} \Big) = \mathrm{sat}_{u^{\min}}^{u^{\max}} \underbrace{(u_I + u_P)}_{u^{cmd}} \qquad (1.6)$$

where the achieved saturated total control signal $u$ is composed of the integral and the proportional feedback terms, $u_I$ and $u_P$, respectively, and $\mathrm{sat}_{u^{\min}}^{u^{\max}}$ represents the static saturation function, defined component-wise, to represent (1.5).

If a component of the commanded control $u^{cmd}$ exceeds its min/max bound then that control channel is saturated at the corresponding limit and the system becomes open-loop with respect to that control input component. During control saturation events, the controller integrator state $e_{yI}$ needs to be kept bounded irrespective of the tracking error dynamics, which in turn can drive the integrator state to become unbounded, that is it would "wind-up".

In order to prevent the integrator from winding up, consider the following anti-windup (AW) modification to the integrator dynamics,

$$\dot{e}_{yI} = y_{reg} - y_{cmd} + v \qquad (1.7)$$

with the AW control input signal $v \in R^m$ to be designed such that the following two conditions are met:

1) The integrator state $e_{yI}$ is uniformly bounded.

2) The closed-loop system remains stable.

This is the integrator AW modification design problem. Clearly, with the AW input $v$ in (1.7), closed-loop tracking performance will be modified and possibly degraded, at the expense of keeping the total control values within the predefined min/max bounds, while retaining closed-loop stability of the system.

Background, motivation, design and relevant discussions for AW design modifications in servo-controllers can be found in [7] and [8]. A majority of these methods may not have a formal basis. However, a lack of control-theoretic fundamentals for these AW algorithms does not preclude their practicality and applicability, which in turn originated through numerous successful applications in industrial systems. In contrast, the developed in this paper AW algorithm does have a control-theoretic basis, but it has not yet been fielded in a realistic setup. Only simulation tests are performed to illustrate key features and potential benefits of the design.

The rest of the paper is organized as follows. The AW design methodology is presented in Section 2. The design is based on the Nagumo Theorem [1] and the theory of Control Barrier Functions (CBFs) [2], [3], [5]. The proposed AW control modification is applicable to servo-controllers with integral feedback. The solution is computed explicitly via a constrained minimization problem [4]. Closed-loop system stability during saturation events is discussed in Section 3. Simulation test data representative of an aircraft pitching motion with a position-limited servo-controller in the loop are analyzed in Section 4. The paper ends with a concise summary of the presented material.

## 2. Integrator AW Design via Control Barrier Functions

Consider the original system (1.1) operating under position-limited servo-controller (1.6), with the modified integrator dynamics (1.7).

$$\dot{e}_{yI} = y_{reg} - y_{cmd} + v$$
$$\dot{x}_p = A_p x_p + B_p \underbrace{\left( \underbrace{\text{sat}_{u^{\min}}^{u^{\max}} \left( -K_I e_{yI} - K_P x_p \right)}_{u^{cmd}} \right)}_{u} \tag{2.1}$$

The AW control modification input $v$ needs to be found such that the integrator state $e_{yI}$ is uniformly bounded forward in time, $\|e_{yI}\| < \infty$, and the closed loop system remains stable.

Towards that end, consider the following minimization problem with double-inequality constraints,

$$\begin{aligned} \text{Minimizaion Cost} &: \|v\|^2 \to \min_v \\ m - \text{Constraints} &: u^{\min} \leq \underbrace{\left( -K_I e_{yI} - K_P x_p \right)}_{u^{cmd}} \leq u^{\max} \end{aligned} \tag{2.2}$$

Define the commanded control deficiency signal,

$$\Delta u^{cmd}(x) = \text{sat}_{u^{\min}}^{u^{\max}} \left( u^{cmd}(x) \right) - u^{cmd}(x) \tag{2.3}$$

and rewrite (2.2) as the minimization problem with $(2m)$ inequality constraints.

$$\begin{aligned} \text{Minimizaion Cost} &: \|v\|^2 \to \min_v \\ \text{Constraints} &: \left[ \Delta u^{cmd}(x) = 0 \right] \Leftrightarrow \left[ g(x) = \begin{pmatrix} g_1(x) = u^{\min} - u^{cmd} \\ g_2(x) = u^{cmd} - u^{\max} \end{pmatrix} \leq 0 \right] \end{aligned} \tag{2.4}$$

By Nagumo's Theorem [1], [2], this constrained optimization problem is equivalent to

$$\text{Minimizaion Cost}: \|v\|^2 \to \min_{v}$$

$$\text{Nagumo's Constraints}: \dot{g}(x)\big|_{g(x)=0} = \begin{pmatrix} \dot{g}_1(x)\big|_{g_1(x)=0} \\ \dot{g}_2(x)\big|_{g_2(x)=0} \end{pmatrix} \leq 0 \tag{2.5}$$

where

$$\dot{g}_k(x) = \frac{\partial g_k(x)}{\partial x}\dot{x} = \frac{\partial g_k(x)}{\partial x}\left(Ax + B\,\text{sat}_{u^{\min}}^{u^{\max}}(u^{cmd})\right), \quad k=1,2 \tag{2.6}$$

is the time derivative of the $k^{th}$ constraint function $g_k(x)$, computed along the trajectories of the extended system (1.3) and evaluated on the boundary set defined by that constraint $g_k(x)=0$.

According to the CBF design [3], the boundary constraints in (2.5) can be replaced by the set constraints,

$$\text{Minimizaion Cost}: \|v\|^2 \to \min_{v}$$

$$\text{CBF Constraints}: G(x,v) = \begin{pmatrix} \dot{g}_1(x) + \alpha_{cbf}\,g_1(x) \\ \dot{g}_2(x) + \alpha_{cbf}\,g_2(x) \end{pmatrix} \leq 0 \tag{2.7}$$

where $\alpha_{cbf} > 0$ is a positive constant parameter. Substituting (2.6) into (2.7), gives

$$G(x,v) = \begin{pmatrix} G_1(x,v) \\ G_2(x,v) \end{pmatrix} = \begin{pmatrix} K_I(e_y + v) + K_P\overbrace{\left(A_p x_p + B_p\,\text{sat}_{u^{\min}}^{u^{\max}}(u^{cmd})\right)}^{\dot{x}_p} + \alpha_{cbf}(u^{\min} - u^{cmd}) \\ -K_I(e_y + v) - K_P\underbrace{\left(A_p x_p + B_p\,\text{sat}_{u^{\min}}^{u^{\max}}(u^{cmd})\right)}_{\dot{x}_p} + \alpha_{cbf}(u^{cmd} - u^{\max}) \end{pmatrix} \tag{2.8}$$

Rewrite (2.8) as,

$$G(x,v) = \begin{pmatrix} I_m \\ -I_m \end{pmatrix} K_I v + \underbrace{\begin{pmatrix} I_m \\ -I_m \end{pmatrix}\left(K_I e_y + K_P \dot{x}_p\right) + \alpha_{cbf}\begin{pmatrix} u^{\min} - u^{cmd} \\ u^{cmd} - u^{\max} \end{pmatrix}}_{\Delta(x)} \tag{2.9}$$

where $\Delta(x)$ denotes the terms in (2.9) that do not explicitly depend on $v$.

$$\Delta(x) = \begin{pmatrix} \Delta_1(x) \\ \Delta_2(x) \end{pmatrix} = \begin{pmatrix} I_m \\ -I_m \end{pmatrix}\left(K_I e_y + K_P\left(A_p x_p + B_p\,\text{sat}_{u^{\min}}^{u^{\max}}(u^{cmd})\right)\right) + \alpha_{cbf}\begin{pmatrix} u^{\min} - u^{cmd} \\ u^{cmd} - u^{\max} \end{pmatrix} \tag{2.10}$$

Then the CBF constraints in (2.7) become,

$$G(x,v) = \begin{pmatrix} G_1(x,v) \\ G_2(x,v) \end{pmatrix} = \begin{pmatrix} I_m \\ -I_m \end{pmatrix} K_I v + \Delta(x) \leq 0 \qquad (2.11)$$

and the resulting constrained minimization problem can be solved analytically via the Lagrange' multiplication coefficients method [4], [5]. Note that since the minimization cost is quadratic in the AW control decision variable $v$ and the CBF constraints (2.11) are linear in $v$, the optimal minimizing solution exists and it is unique.

For the constrained minimization problem (2.7)-(2.11), consider the Lagrangian function,

$$L(x,v,\lambda) = \|v\|^2 + \lambda^T G(x,v) = v^T v + \lambda^T \left( \begin{pmatrix} I_m \\ -I_m \end{pmatrix} K_I v + \Delta(x) \right) \qquad (2.12)$$

with the Lagrange multiplier vector-coefficient

$$\lambda = \begin{pmatrix} \lambda_1 \\ \lambda_2 \end{pmatrix} \in R^{2m}, \quad \lambda_k \in R^m, \quad k = 1, 2 \qquad (2.13)$$

The minimization function is convex and the linear inequality constraints are continuously differentiable with respect to the AW control decision variable $v$. Consequently, Karush-Kuhn-Tucker (KKT) conditions for optimality are applicable, for any $x \in R^n$ and $v \in R^m$, [4].

$$\begin{aligned} &\text{Stationarity}: \frac{\partial L(x,v,\lambda)}{\partial v} = 0 \\ &\text{Primal Feasibility}: G(x,v) \leq 0 \\ &\text{Dual Feasibility}: \lambda \geq 0 \\ &\text{Complimentary Slackness}: \lambda^T G(x,v) = 0 \end{aligned} \qquad (2.14)$$

Solving the KKT stationarity condition,

$$\frac{\partial L(x,v,\lambda)}{\partial v} = \frac{\partial \left( v^T v + \lambda^T \left( \begin{pmatrix} I_m \\ -I_m \end{pmatrix} K_I v + \Delta(x) \right) \right)}{\partial v} = 2v + K_I^T \begin{pmatrix} I_m & -I_m \end{pmatrix} \lambda = 0 \qquad (2.15)$$

for the optimal AW control decision policy $v$, gives

$$v = -0.5 K_I^T \begin{pmatrix} I_m & -I_m \end{pmatrix} \lambda = -0.5 K_I^T (\lambda_1 - \lambda_2) \qquad (2.16)$$

In order to solve for $\lambda$, the CBF inequality constraints (2.11) need to be evaluated along with (2.16) at the constraints boundary.

$$\begin{aligned}
\left[G_1(x,v)=0,\ \lambda_1\geq 0\right] &\Rightarrow \left[K_I\left(e_y-0.5K_I^T(\lambda_1-\lambda_2)\right)+K_P\dot{x}_p+\alpha_{cbf}\left(u^{\min}-u^{cmd}\right)=0\right] \\
\left[G_2(x,v)=0,\ \lambda_2\geq 0\right] &\Rightarrow \left[-K_I\left(e_y-0.5K_I^T(\lambda_1-\lambda_2)\right)-K_P\dot{x}_p+\alpha_{cbf}\left(u^{cmd}-u^{\max}\right)=0\right]
\end{aligned} \quad (2.17)$$

The first and the second set of $m$-equations in (2.17) are mutually exclusive.

$$\begin{aligned}
\left[\lambda_1\geq 0\right]&\Leftrightarrow\left[\lambda_2=0\right] \\
\left[\lambda_1=0\right]&\Leftrightarrow\left[\lambda_2\geq 0\right]
\end{aligned} \quad (2.18)$$

Consequently, (2.17) reduces to

$$\begin{aligned}
\left[G_1(x,v)=0,\ \lambda_1\geq 0\right] &\Rightarrow \left[K_I\left(e_y-0.5K_I^T\lambda_1\right)+K_P\dot{x}_p+\alpha_{cbf}\left(u^{\min}-u^{cmd}\right)=0\right] \\
\left[G_2(x,v)=0,\ \lambda_2\geq 0\right] &\Rightarrow \left[-K_I\left(e_y+0.5K_I^T\lambda_2\right)-K_P\dot{x}_p+\alpha_{cbf}\left(u^{cmd}-u^{\max}\right)=0\right]
\end{aligned} \quad (2.19)$$

Solving each set of equations for the corresponding $\lambda_k$, gives the two Lagrange vector coefficients and the resulting optimal integrator AW control strategy.

$$\left.\begin{aligned}
\lambda_1 &= 2\max\left[0_{m\times 1},\ \underbrace{\left(K_I K_I^T\right)^{-1}\overbrace{\left(K_I e_y+K_P\dot{x}_p+\alpha_{cbf}\underbrace{\left(u^{\min}-u^{cmd}\right)}_{g_1(x)}\right)}^{\Delta_1(x)}}\right] \\
\lambda_2 &= 2\max\left[0_{m\times 1},\ \underbrace{\left(K_I K_I^T\right)^{-1}\left(-K_I e_y-K_P\dot{x}_p+\alpha_{cbf}\underbrace{\left(u^{cmd}-u^{\max}\right)}_{g_2(x)}\right)}_{\Delta_2(x)}\right]
\end{aligned}\right\} \Rightarrow \boxed{v=-0.5K_I^T(\lambda_1-\lambda_2)} \quad (2.20)$$

The derived AW control signal (2.20) represents a linear state feedback control policy and so, standard linear control robustness analysis methods can be applied to determine the corresponding closed-loop system stability, as well as relative stability robustness characteristics, such as gain and phase margins.

For clarity, the AW design equations are summarized in the table below.

| Open-loop MIMO plant dynamics (1.1) | $\dot{x}_p(t)=A_p x_p(t)+B_p u(t)$ |
|---|---|
| Regulated output (1.1) | $y_{reg} = C_{p\,reg}\, x_p(t) + D_{p\,reg}\, u(t)$ |
| Unlimited integrator state dynamics (1.2) | $\dot{e}_{yI} = y_{reg} - y_{cmd} = e_y$ |
| Extended open-loop system (1.3) | $\underbrace{\begin{pmatrix} \dot{e}_{yI} \\ \dot{x}_p \end{pmatrix}}_{\dot{x}} = \underbrace{\begin{pmatrix} 0_{m\times m} & C_{p\,reg} \\ 0_{n_p \times m} & A_p \end{pmatrix}}_{A} \underbrace{\begin{pmatrix} e_{yI} \\ x_p \end{pmatrix}}_{x} + \underbrace{\begin{pmatrix} D_{p\,reg} \\ B_p \end{pmatrix}}_{B} u + \underbrace{\begin{pmatrix} -I_m \\ 0 \end{pmatrix}}_{B_{cmd}} y_{cmd}$ |
| Servo-controller with position limits (1.6) | $u = \mathrm{sat}_{u^{\min}}^{u^{\max}}\left( \underbrace{-K_I\, e_{yI}}_{u_I}\ \underbrace{-K_P\, x_p}_{u_P} \right) = \mathrm{sat}_{u^{\min}}^{u^{\max}}\underbrace{(u_I + u_P)}_{u^{cmd}}$ |
| Commanded control input | $u^{cmd} = \underbrace{-K_I\, e_{yI}}_{u_I}\ \underbrace{-K_P\, x_p}_{u_P}$ |
| Closed-loop system with the servo-controller and the AW modified integrator (2.1) | $\dot{e}_{yI} = e_y + v$ <br> $\dot{x}_p = A_p\, x_p + B_p\, \mathrm{sat}_{u^{\min}}^{u^{\max}}(u^{cmd})$ |
| Control constraints (2.4) | $g(x) = \begin{pmatrix} g_1(x) = u^{\min} - u^{cmd} \\ g_2(x) = u^{cmd} - u^{\max} \end{pmatrix} \le 0$ |
| State dependent constraints function (2.10) | $\Delta(x) = \begin{pmatrix} \Delta_1(x) \\ \Delta_2(x) \end{pmatrix} = \begin{pmatrix} K_I\, e_y + K_P\, \dot{x}_p + \alpha_{cbf}\, g_1(x) \\ -K_I\, e_y - K_P\, \dot{x}_p + \alpha_{cbf}\, g_2(x) \end{pmatrix}$ |
| Lagrange multiplier vector coefficients (2.20) | $\lambda_1 = 2\max\left[ 0_{m\times 1},\ (K_I K_I^T)^{-1} \Delta_1(x) \right]$ <br> $\lambda_2 = 2\max\left[ 0_{m\times 1},\ (K_I K_I^T)^{-1} \Delta_2(x) \right]$ |
| AW control solution (2.20) | $v = -0.5\, K_I^T (\lambda_1 - \lambda_2)$ |

**Table 1  Integrator AW Control Design Summary**

By design, the AW controller (2.20) enforces uniform boundedness for the integrator state of the servo-controller dynamics (2.1).

In the next section, it is shown that for open-loop stable systems, closed-loop stability is preserved, while preventing the integrator windup.

3.  **Closed-Loop Stability Analysis**

Consider the original system (1.1) with the servo-controller (2.1) and the AW modification (2.20) in the loop.

$$\dot{e}_{yI} = e_y - 0.5 K_I^T (\lambda_1 - \lambda_2)$$
$$\dot{x}_p = A_p x_p + B_p \underbrace{\mathrm{sat}_{u^{\min}}^{u^{\max}} \underbrace{(-K_I e_{yI} - K_P x_p)}_{u^{cmd}}}_{u} \qquad (3.1)$$

Using (2.10) and (2.20), the integrator dynamics in (3.1) can be expressed as,

$$\dot{e}_{yI} = e_y - 0.5 K_I^T (\lambda_1 - \lambda_2)$$
$$= \begin{cases} e_y - K_I^T (K_I K_I^T)^{-1} (K_I e_y + K_P \dot{x}_p + \alpha_{cbf} g_1(x)), & \text{if } u^{cmd} \leq u^{\min} \\ e_y + K_I^T (K_I K_I^T)^{-1} (-K_I e_y - K_P \dot{x}_p + \alpha_{cbf} g_2(x)), & \text{if } u^{cmd} \geq u^{\max} \\ e_y, & \text{otherwise} \end{cases}$$
$$= \begin{cases} -K_I^{-1} (K_P (A_p x_p + B_p u^{\min}) + \alpha_{cbf} (u^{\min} - u^{cmd})), & \text{if } u^{cmd} \leq u^{\min} \\ K_I^{-1} (-K_P (A_p x_p + B_p u^{\max}) + \alpha_{cbf} (u^{cmd} - u^{\max})), & \text{if } u^{cmd} \geq u^{\max} \\ e_y, & \text{otherwise} \end{cases} \qquad (3.2)$$
$$= \begin{cases} -\alpha_{cbf} e_{yI} - K_I^{-1} (K_P (A_p x_p + B_p u^{\min}) + \alpha_{cbf} (u^{\min} + K_P x_p)), & \text{if } u^{cmd} \leq u^{\min} \\ -\alpha_{cbf} e_{yI} + K_I^{-1} (-K_P (A_p x_p + B_p u^{\max}) + \alpha_{cbf} (-K_P x_p - u^{\max})), & \text{if } u^{cmd} \geq u^{\max} \\ e_y, & \text{otherwise} \end{cases}$$

Rewrite (3.2) as

$$\dot{e}_{yI} = \begin{cases} -\alpha_{cbf} e_{yI} - K_I^{-1} K_P (A_p + \alpha_{cbf} I_{n_p}) x_p - K_I^{-1} (\alpha_{cbf} + K_P B_p) u^{\min}, & \text{if } u^{cmd} \leq u^{\min} \\ -\alpha_{cbf} e_{yI} - K_I^{-1} K_P (A_p + \alpha_{cbf} I_{n_p}) x_p - K_I^{-1} (\alpha_{cbf} + K_P B_p) u^{\max}, & \text{if } u^{cmd} \geq u^{\max} \\ e_y, & \text{otherwise} \end{cases}$$
$$= \begin{cases} -\alpha_{cbf} e_{yI} - K_I^{-1} K_P (A_p + \alpha_{cbf} I_{n_p}) x_p - K_I^{-1} (\alpha_{cbf} + K_P B_p) \mathrm{sat}_{u^{\min}}^{u^{\max}} (u^{cmd}), & \text{if } \begin{pmatrix} (u^{cmd} \leq u^{\min}) \\ \vee \\ (u^{cmd} \geq u^{\max}) \end{pmatrix} \\ e_y, & \text{otherwise} \end{cases} \qquad (3.3)$$

From (3.3) it follows that during control saturation events, the closed-loop system dynamics (3.1) can be written as,

$$\dot{e}_{yI} = -\alpha_{cbf}\, e_{yI} - K_I^{-1} K_P \left( A_p + \alpha_{cbf}\, I_{n_p} \right) x_p - K_I^{-1} \left( \alpha_{cbf} + K_P B_p \right) \text{sat}_{u^{min}}^{u^{max}} \left( u^{cmd} \right)$$
$$\dot{x}_p = A_p\, x_p + B_p\, \text{sat}_{u^{min}}^{u^{max}} \left( u^{cmd} \right) \tag{3.4}$$

or equivalently, in matrix form.

$$\begin{pmatrix} \dot{e}_{yI} \\ \dot{x}_p \end{pmatrix} = \underbrace{\begin{pmatrix} -\alpha_{cbf}\, I_m & -K_I^{-1} K_P \left( A_p + \alpha_{cbf}\, I_n \right) \\ 0_{n_p \times m} & A_p \end{pmatrix}}_{\tilde{A}_{cl}} \begin{pmatrix} e_{yI} \\ x_p \end{pmatrix} + \underbrace{\begin{pmatrix} -K_I^{-1} \left( \alpha_{cbf} + K_P B_p \right) \\ B_p \end{pmatrix} \text{sat}_{u^{min}}^{u^{max}} \left( u^{cmd} \right)}_{\text{Constant Term: } C_0} \tag{3.5}$$

Clearly, the eigenvalues (EV-s) of this system are

$$\lambda\left(\tilde{A}_{cl}\right) = \underbrace{\left(-\alpha_{cbf}\right)}_{m\ EV-s} \cup \underbrace{\lambda\left(A_p\right)}_{n_p\ EV-s} \tag{3.6}$$

and consequently, the closed-loop system retains stability if and when the control input components reach their min/max limits. Otherwise, the closed-loop system is stable by the design.

## 4. Flight Control Design and Simulation Trade Study

Consider the aircraft short-period dynamics [6].

$$\underbrace{\begin{pmatrix} \dot{\alpha} \\ \dot{q} \end{pmatrix}}_{\dot{x}_p} = \underbrace{\begin{pmatrix} \dfrac{Z_\alpha}{V_0} & 1 + \dfrac{Z_q}{V_0} \\ M_\alpha & M_q \end{pmatrix}}_{A_p} \underbrace{\begin{pmatrix} \alpha \\ q \end{pmatrix}}_{x_p} + \underbrace{\begin{pmatrix} \dfrac{Z_{\delta_e}}{V_0} \\ M_{\delta_e} \end{pmatrix}}_{B_p} \underbrace{\delta_e}_{u}$$
$$\underbrace{\alpha}_{y_{reg}} = \underbrace{(1\ \ 0)}_{C_{p\,reg}} x_p + \underbrace{0}_{D_{p\,reg}} u \tag{4.1}$$

where the elevator deflection $\delta_e$ represents the system control input in radians, the regulated output $\alpha$ is the Angle of Attack (AOA) in radians and $q$ is the pitch rate in radians/sec. The aircraft open-loop linear model data are defined below.

$$A_p = \begin{pmatrix} -2.241 & 0.9897 \\ -4.474 & -0.9024 \end{pmatrix},\quad B_p = \begin{pmatrix} -0.23307 \\ -4.5926 \end{pmatrix}$$
$$C_{p\,reg} = (1\ \ 0) \quad,\quad D_{p\,reg} = 0 \tag{4.2}$$

These open-loop data are representative of a piloted tactical aircraft, trimmed wings-level at high speed flight conditions.

The open-loop system is stable and therefore the AW control design methodology (2.20) is applicable. The overall control goal is to design an optimal state feedback servo-controller (1.6), with the AW modification (2.20) to prevent the controller integrator from winding up, while retaining closed-loop system stability, with possibly degraded command tracking performance, during control saturation events.

For this example and in order to demonstrate key features of the AW controller (2.20), the system control position limits are defined relatively small, $(\pm 10 \deg)$.

The LQR method is employed [6] to design a robust baseline PI servo-controller in the form of (1.4), using the extended open-loop system dynamics (1.3). With the selected optimal weights,

$$Q = \begin{pmatrix} 20 & 0 & 0 \\ 0 & 0 & 0 \\ 0 & 0 & 0.2 \end{pmatrix}, \quad R = 1$$

the baseline LQR PI servo-controller gains (1.4) are computed.

$$K_x = \left( \underbrace{-4.4721}_{K_I} \quad \underbrace{-1.0369 \quad -0.58504}_{K_P} \right)$$

Figure 1 shows the system loop gain (Lu LQR) computed at the control input break point and the closed-loop system step-response.

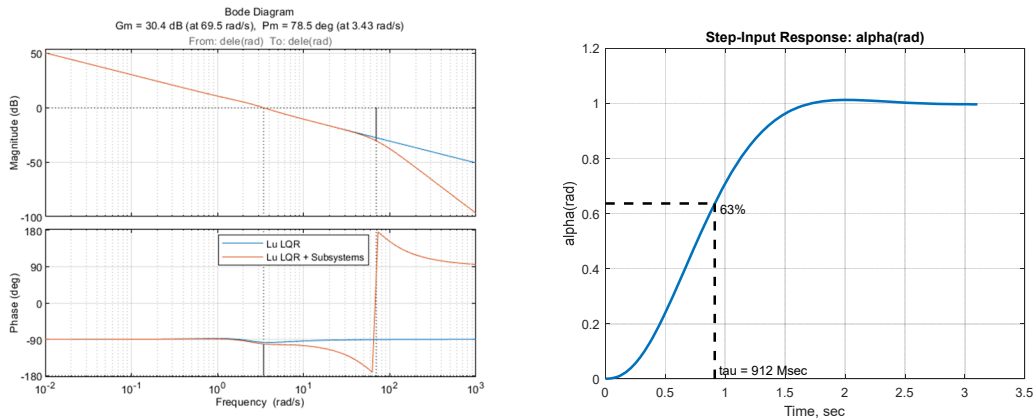

**Figure 1  Baseline Position-Unlimited State-Feedback Servo-Controller Analysis Data**

The servo-control design has sufficient stability margins and adequate step-response characteristics. In this case, "subsystems" used for analysis (but not for the design) include a second order control actuator with the natural frequency of $70(rps)$ and the damping of $0.7$.

Figure 2 shows closed-loop simulation data for a 10 deg command doublet maneuver, with the baseline position-unlimited servo-controller (1.4) in the loop and without the AW modification (2.20).

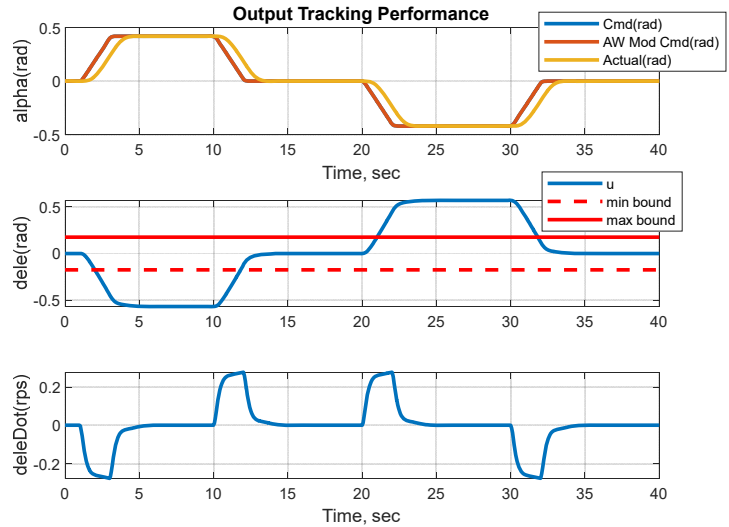

**Figure 2  Closed-Loop Simulation with Position-Unlimited Baseline Servo-Controller**

The command tracking performance (first subplot) is smooth but the required control values (second subplot) clearly exceed the min/max actuator position limits. The control rate (third subplot) is also shown in order to verify reasonable actuation demands.

Note the legend in the first subplot with the regulated output data. Shown there are: a) The external command, b) The AW command modification $(y_{cmd} + v)$, and c) The actual system response. In this example, the first two signals are identical since the AW controller is Off. The time lag between the command (blue) and the system response (yellow) is as expected and per design. Decreasing the lag can be accomplished by adding a command-feedforward term into the controller and/or by an increasing the controller bandwidth, which in turn would require higher feedback gains and a potential degradation of the system stability margins. Often for a piloted aircraft, a command-feedforward term would be added to the servo-controller in order to improve flying qualities without margin degradations.

Figure 3 shows closed-loop simulation data with the position-limited PI servo-controller but without the integrator AW modification.

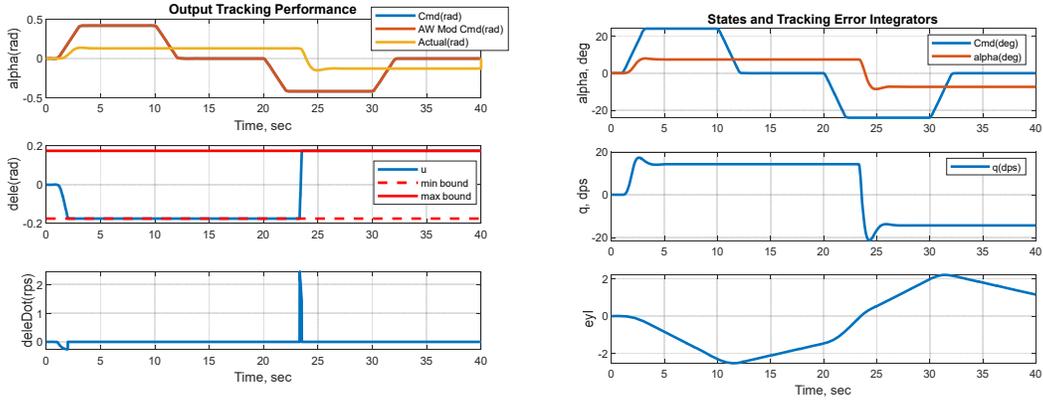

**Figure 3  Closed-Loop Simulation with Position-Limited Servo-Controller without AW Modification**

As seen from the data, the closed-loop system performance (left and right first subplots) is significantly degraded. The time-delay in the system response is very large and would be totally unacceptable in a realistic flight control application. The significantly delayed response of the regulated output is primarily due to winding up of the integrator dynamics (right third subplot) during control saturation.

Turning on the AW control modification (2.20), removes the undesirable time-lag in the system response and as such, improves overall closed-loop system performance, (Figure 4).

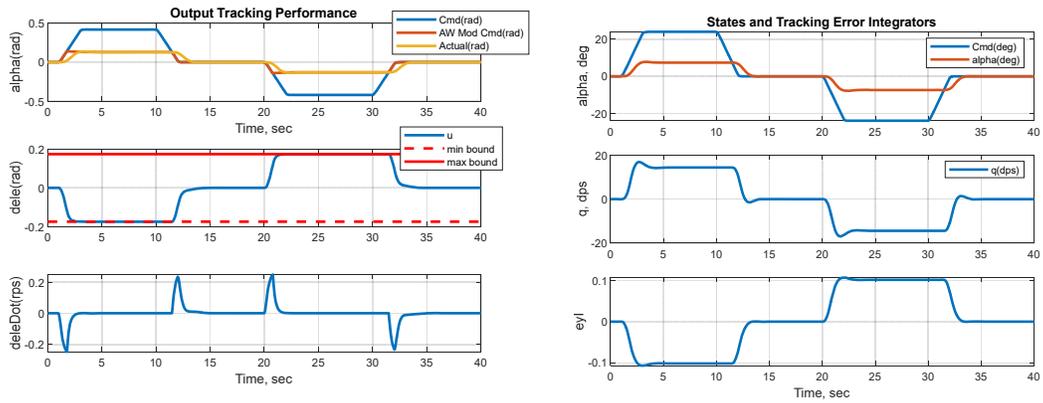

**Figure 4  Closed-Loop Simulation with Position-Limited Servo-Controller and AW Modification**

With the AW control modification turned on, the integrator state dynamics (right third plot) are managed such that the regulated output (left first plot) tracks the same modified command value as in Figure 3 but the system response time-delay time delay is restored back to the baseline controller level (Figure 1), resulting in a degraded (due to control saturation) but acceptable (achievable) regulated output tracking performance.

In this case, the AW positive constant parameter $\alpha_{cbf}$ for the CBF constraints (2.7) is defined as

$$\alpha_{cbf} = \|K_I\| = 4.4721$$

Figure 5 (left plots) shows the two Lagrange multiplier coefficients, $\lambda_1$ and $\lambda_2$, from (2.20), along with the two corresponding min/max CBF constraints, $G_1(x,v)$ and $G_2(x,v)$, as defined in (2.8).

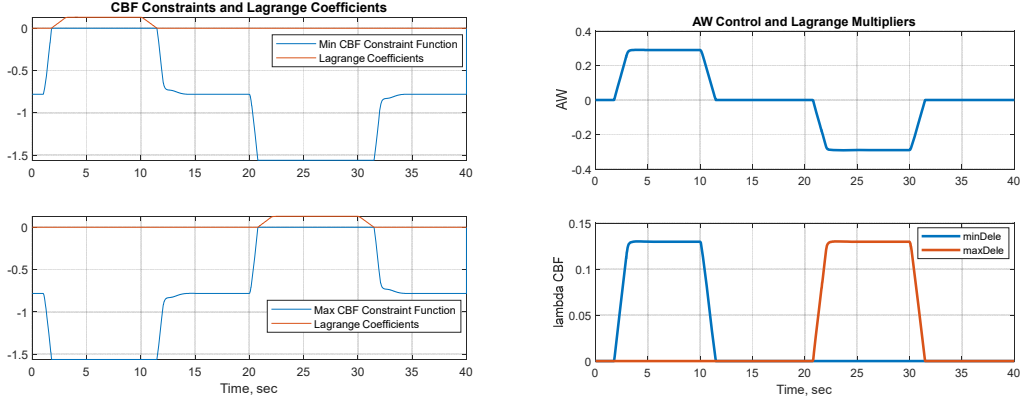

**Figure 5  Closed-Loop Simulation with Position-Limited Servo-Controller and AW Modification**

The Lagrange coefficients (left plots, red lines, right second subplot blue and red lines) become non-zero and strictly positive only when the CBF constraints (left subplots, blue lines) are violated. This is per design and as expected. The resulting AW control modification signal (2.20) (right first subplot) is smooth.

The positive constant parameter $\alpha_{cbf}$ is the primary tuning knob for the AW design. Large values of $\alpha_{cbf}$ result in smaller/faster dynamics for the Lagrange coefficients. That in turn, gives a lesser overshoot for the commanded control signal over its prescribed min/max position limits. These observations are substantiated by the stability analysis in Section 3, whereby as seen from (3.5), large values of $\alpha_{cbf}$ yield a smaller time constant for the integrator dynamics, resulting in a faster convergence of the AW-modified integrator state.

In order to assess sensitivity of the AW design, an unknown (to the controller) external disturbance $d$ is added to the system dynamics to emulate an environmental uncertainty, such as wind/gust.

$$\underbrace{\begin{pmatrix} \dot{\alpha} \\ \dot{q} \end{pmatrix}}_{\dot{x}_p} = \underbrace{\begin{pmatrix} \dfrac{Z_\alpha}{V_0} & 1 + \dfrac{Z_q}{V_0} \\ M_\alpha & M_q \end{pmatrix}}_{A_p} \underbrace{\begin{pmatrix} \alpha \\ q \end{pmatrix}}_{x_p} + \underbrace{\begin{pmatrix} \dfrac{Z_{\delta_e}}{V_0} \\ M_{\delta_e} \end{pmatrix}}_{B_p} \underbrace{\delta_e}_{u} + \begin{pmatrix} 1 \\ 0 \end{pmatrix} d$$

By definition, $d$ represents an unknown incremental change in AOA rate of change. For simulation testing, the disturbance data are selected as a sinusoidal, $d = 4\dfrac{\pi}{180}\sin(2t)$. Figure 6 shows the data.

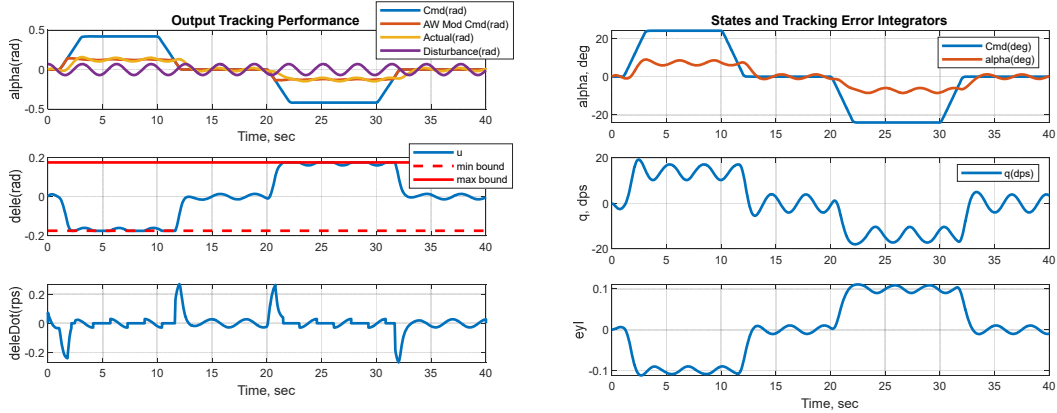

**Figure 6 Closed-Loop Simulation with Position-Limited Servo-Controller, AW Modification and Unknown Disturbance**

Disturbance rejection and overall system robustness properties are clearly visible from the test data. Other simulation tests performed with several low-to-high frequency set of disturbances confirmed the robustness of the design. Control-theoretic justifications of these observations are currently under investigation and will be reported elsewhere.

## 5. Conclusions

In this paper, a formal AW modification control design is derived for MIMO LTI open-loop stable systems with servo-controllers operating within predefined control position min/max bounds. The AW design is based on the CBF methods [3], [5]. Simulation tests are presented to illustrate the design steps and key features.

## References

[1] M. Nagumo, "Über die Lage der Integralkurven gewöhnlicher differentialgleichungen," in Proc. Physico-Math. Soc. Jpn., in 3rd Series, vol. 24, Jan. 1942, pp. 551–559.

[2] F. Blanchini, "Set invariance in control," Automatica, vol. 35, no. 11, pp. 1747–1767, 1999, https://doi.org/10.1016/S0005-1098(99)00113-2

[3] A.D. Ames, X. Xu, J.W. Grizzle, P. Tabuada, "Control barrier function based quadratic programs for safety critical systems," IEEE Trans. Autom. Control, vol. 62, no. 8, pp. 3861–3876, Aug. 2017, 10.1109/TAC.2016.2638961

[4] S. Boyd, L. Vandenberghe, *Convex Optimization*. Cambridge, U.K., Cambridge Univ. Press, 2004, https://doi.org/10.1017/cbo9780511804441

[5] A. Alan, A.J. Taylor, C.R. He, A.D. Ames, G. Orosz, "Control barrier function and input-to-state safety with application to automated vehicles," IEEE Trans. Control Systems Technology, vol. 31, no. 6, pp. 2744–2759, Nov. 2023, https://doi.org/10.48550/arXiv.2206.03568

[6] E. Lavretsky, K.A. Wise, *Robust and Adaptive Control with Aerospace Applications*, Second Edition, Advanced Textbooks in Control and Signal Processing, Springer Nature Switzerland AG, ISBN 978-3-031-38313-7 (print), ISBN 978-3-031-38314-4 (eBook), 2024, https://doi.org/10.1007/978-3-031-38314-4.

[7] K.J. Åström, R.M. Murray, *Feedback systems: an introduction for scientists and engineers*, Princeton University Press, 2008, https://doi.org/10.1515/9781400828739

[8] S. Tarbouriech, M.C. Turner, "Anti-windup design: an overview of some recent advances and open problems", IET Control Theory and Application, vol. 3, no. 1, pp. 1-19, 2009, 10.1049/iet-cta:20070435